\documentclass[fleqn,usenatbib]{mnras}

\usepackage{newtxtext,newtxmath}

\usepackage[T1]{fontenc}

\DeclareRobustCommand{\VAN}[3]{#2}
\let\VANthebibliography\thebibliography
\def\thebibliography{\DeclareRobustCommand{\VAN}[3]{##3}\VANthebibliography}

\usepackage{graphicx}	
\usepackage{subfig}
\usepackage{stfloats}
\usepackage{amsmath}	

\usepackage{amssymb}	
\usepackage{CJKutf8}
\usepackage{tabularx}

\title[Modelling of large-scale horse-shoe-like filament]{Modelling the magnetic structure of a large-scale horse-shoe-like filament in a decaying and diffuse active region}

\author[Kang et al.]{
Kaifeng Kang,$^{1,2,3}$
Yang Guo,$^{4}$\thanks{E-mail: \url{guoyang@nju.edu.cn}}
Ilia I. Roussev,$^{5}$
Rony Keppens$^{6}$
and Jun Lin$^{1,2,3}$\thanks{E-mail: \url{jlin@ynao.ac.cn}}
\\
$^{1}$Yunnan Observatories, Chinese Academy of Sciences, PO Box 110, Kunming, Yunnan 650216, China\\
$^{2}$University of Chinese Academy of Sciences, Beijing 100049, China\\
$^{3}$Center for Astronomical Mega-Science, Chinese Academy of Sciences, Beijing 100101, China\\
$^{4}$School of Astronomy and Space Science and Key Laboratory for Modern Astronomy and Astrophysics, Nanjing University, Nanjing,  Jiangsu 210023, China\\
$^{5}$Space Sciences Laboratory, University of California Berkeley, 7 Gauss Way, Berkeley, CA 94720, USA\\
$^{6}$Centre for Mathematical Plasma Astrophysics, Department of Mathematics, KU Leuven, Celestijnenlaan 200B, B-3001 Leuven, Belgium\\
}

\date{Accepted XXX. Received YYY; in original form ZZZ}

\pubyear{2022}

\begin{document}
\label{firstpage}
\pagerange{\pageref{firstpage}--\pageref{lastpage}}
\maketitle

\begin{abstract}
A large-scale, horse-shoe-like filament was investigated and the magnetic field around it was reconstructed. This is an intermediate filament (IF) that appeared on the solar disk for the first time at 02:00 UT on 2015 November 7, and took 8 days to move to the central median on the solar disk. The active region AR 12452 around which the filament occurred was diffuse so that the magnetic field nearby was weak, the average field strength is 106 G. Therefore, the existing approaches to extrapolating the coronal magnetic field and to constructing the filament configuration in the region with strong background field do not work well here. On the basis of the regularized Biot-Savart laws method, we successfully constructed a data-constrained, non-linear force-free field configuration for this IF observed on 2015 November 14. The overall IF configuration obtained in this way matches well the morphology suggested by a 304~\AA \ image taken by the Atmospheric Imaging Assembly on board Solar Dynamics Observatory. Magnetic dips in the configuration were coincident in space with the H$\alpha$ features of the filament,   which is lower in altitude than the features seen in 304~\AA. This suggests that the cold plasma fills the lower part of the filament, and hot plasma is situated in the higher region. A quasi-separatrix layer wraps the filament, and both the magnetic field and the electric current are stronger near the inner edge of the filament.

\end{abstract}

\begin{keywords}
Sun:filaments,prominences--Sun:magnetic fields--Sun:activity-- Sun:photosphere--Sun:chromosphere--Sun:corona
\end{keywords}



\section{Introduction}

A solar filament is a magnetic structure that floats in the corona and includes partially ionized plasma, which is about 100 times cooler and denser than their coronal surroundings \citep{2010SSRv..151..243L}. It is  described as filament when seen against the solar disk in absorption and also  known as prominence when observed above the solar limb as bright features against the dark background   \citep{2005SoPh..227..283L}. Filaments can be found in regions of weak background field (quiescent filaments, QF), in the center of active regions (active filaments, AF) and at the border of active regions (intermediate filaments, IF; \citealt{1998ASPC..150...23E}). A filament mainly consists of three typical sub-structures: a spine, which runs horizontally along the filament; some barbs with lateral structures that extend down from the spine to the chromosphere below; and, two extreme ends where the spine anchors in \citep{1998SoPh..182..107M,2010SSRv..151..333M}. These three components of filaments always differ in appearance in QF, in AF, and in IF  \citep{2008ASPC..383..235L}. For QF case, recent full three-dimensional (3D)  magnetohydrodynamic modelling of \citet{2022NatAs.tmp..153J} on their in-situ formation process could resolve the dichotomy in their filament/prominence appearance, with vertical structuring caused by the magnetic Rayleigh-Taylor instability process.

Solar filaments have attracted considerable attention both from observations and theories ever since their first observations because their eruption can lead to the coronal mass ejection (CME), which is the main driver  of the disastrous space weather  \citep{2015SoPh..290.3457S,2018SSRv..214...46G}.
However, the reason why the cool and dense filament can exist in the hot and tenuous corona is still an open question. Magnetic fields are considered to be central to maintaining filament existence in the coronal surroundings \citep{2010SSRv..151..333M}. Therefore, measuring the magnetic field of filaments to obtain their 3D magnetic structure is  key to better understand their structure, evolution and eruption. Unfortunately, the magnetic filed in the upper solar atmosphere is very difficult to measure directly. Until recently, only on the photosphere can magnetic field be routinely measured with better accuracy with a few exceptions as discussed by \citet{2007ASPC..368..291L},  \citet{2013ApJ...777..108S,2014AA...569A..85S}, and \citet{2016ApJ...818...31L,2016ApJ...826..164L}. These authors got full vector magnetic field maps within prominences with spectro-polarimetric data taken by the French telescope, T\'{e}lescope H\'{e}liographique pour 1'Etude du Magn\'{e}tisme et des Instabilit\'{e}s Solaires (THEMIS) in the He D3 line using a spectral inversion code. In most cases, the 3D coronal magnetic field is constructed by magnetic field extrapolation models \citep{1988A&A...198..293S,1990ApJ...352..343L,1992ApJ...399..300L,1997ApJ...486..534M,1997SoPh..174..129A,2000ApJ...540.1150W,2003NPGeo..10..313W}, magnetohydrostatic models \citep{1985ApJ...293...31L,2000A&A...356..735P,2017ApJS..229...18W,2020A&A...640A.103Z}, and magnetohydrodynamic models \citep{2000PhRvL..84.1196A,2014Natur.514..465A,2016ApJ...823...22X}.

The first numerical model for the 3D coronal magnetic field of filaments was developed by \citet{1998A&A...329.1125A} and \citet{1998A&A...335..309A} by extrapolating a photospheric line of sight (LOS) magnetogram into the corona, assuming linear force-free field (LFFF). Since then, ever more sophisticated 3D  models of filaments were also developed by  \citet{1999A&A...342..867A}, \citet{2000ApJ...543..447A},  \citet{2003A&A...402..769A} and \citet{2008SoPh..248...29D}, who  extrapolated photospheric magnetograms into the corona by assuming a linear magnetohydrostatic (LMHS) field. Although these models successfully explained some observational features of filaments, neither assumptions of LFFF and LMHS about coronal magnetic field is realistic. Therefore, more realistic nonlinear force-free field (NLFFF) models were developed for filaments by several authors.

For example, \citet{2002A&A...392.1119R} and \citet{2004A&A...425..345R} constructed NLFFF models for AF by extrapolating observed photospheric vector magnetograms into the corona. Similar case studies can be found in \citet{2016ApJ...818..148L} and \citet{2018ApJ...855L..16J}. \citet{2009SoPh..260..321M} developed another NLFFF model to study how a single bipole polarity advected towards the main body of a filament influences the structure of the filament dips where the magnetic field lines are locally horizontal and curved upward. \citet{2014ApJ...786L..16J} constructed an NLFFF model for a large-scale IF based  on the  photospheric vector magnetogram without any other observational constraint or prerequisite. It is notable that almost all NLFFF models perform well in constructing magnetic structures of small-scale AF and fail in large-scale IF and QF except the case presented in \citet{2014ApJ...786L..16J}, where the involved active region is still very compact even though the coronal magnetic field containing a magnetic flux rope (MFR) was recovered for a large-scale IF. Constructing 3D magnetic models for  QF or large-scale IF involved with large-scale decaying or diffuse active regions is still challenging.

A very creative NLFFF model, which successfully constructs 3D magnetic field for any kinds of QF or IF involved with an MFR, was developed by \citet{2004ApJ...612..519V}. This method is known as the MFR insertion method, which needs to select a path along the polarity inversion line (PIL) where the longitudinal component of the magnetic field changes sign, and to evaluate an ambient potential magnetic field according to the LOS magnetogram. Then, a field-free cavity is created along the selected path and a flux rope is inserted into the cavity by imposing axial and azimuthal magnetic fluxes. Finally, the  magnetic field in such a configuration is numerically relaxed via the magneto-frictional approach developed by \citet{2000ApJ...539..983V} to an NLFFF state with line-tied conditions on the bottom boundary, which is located on the photospheric surface. The MFR insertion method is improved later by \citet{2008ApJ...672.1209B}.

Comparing with the old version, the new method selects the path where an  MFR is inserted according to the location of a filament observed in H$\alpha$ instead of the PIL in the magnetogram. In addition, the model is further constrained by observations of coronal loops in the overlying corona in order to find the solution that best fits the observed features of a filament. This method has been employed to construct 3D magnetic models for observed QF, sigmoidal active regions, and solar polar crown prominence \citep{2009ApJ...691..105S,2009ApJ...703.1766S,2012ApJ...744...78S,2012ApJ...757..168S}.  A magnetic model constructed by the MFR insertion method was also used as the initial condition for MHD simulations of a filament eruption \citep{2013ApJ...779..129K, 2018ApJ...856...75T}. More recently, \citet{2020A&A...637A...3M} constructed a 3D NLFFF model for a prominence by the method, and directly compared for the first time the magnetic field strength deduced from the model with the corresponding results derived from spectro-polarimetric observations of the prominence. A disadvantage of the MFR insertion method relates to the fact that values of the axial flux and the poloidal flux of the inserted flux rope need to be adjusted repeatedly to find the solution that best fits the observed coronal structure. \citet{2014ApJ...790..163T} later addressed this issue in a new model that was developed via the so-called MFR embedding method.

Different from the MFR insertion method, the magnetic vector potential of the flux-rope in the MFR embedding method is purely constructed by analytical form,  on the basis of the property of the ambient configuration \citep{2014ApJ...790..163T}. 
However, the embedded MFR \citep{2014ApJ...790..163T} was restricted to the magnetic configuration of a toroidal-arc shape. Hence, it is difficult to model configurations that reside above a highly elongated or curved PIL via the MFR embedding approach. A breakthrough has been made by \citet{2018ApJ...852L..21T} who removed this geometric limitation of the embedded MFR and deduced the regularized Biot-Savart laws (RBSL), which could model an MFR of arbitrary shape.

In the approach of \citet{2014ApJ...790..163T}, the MFR vector potential is superposed to that of the ambient field instead of creating a field-free cavity where values of the axial flux and poloidal flux of the MFR are changed manually. Comparing with the insertion method, the embedding method has the advantage of  better controlling the  characteristics of the MFR in equilibrium. This is because it defines a force-free flux rope  approximately in equilibrium by using the ambient potential to estimate the axial and the azimuthal magnetic fluxes. Nevertheless, a drawback of the method is that it generally leads to a modification of the observed  photospheric magnetogram in the modelled magnetic configuration by superimposing a flux-rope field on the magnetogram.

\citet{2018ApJ...852L..21T} figured out a way to solve this problem by using a technique similar to that of \citet{2004ApJ...612..519V}. First, they constructed an MFR according to the RBSL method, and extracted the longitudinal field of the MFR inside the footpoint area on the photosphere. Second, the extracted longitudinal field was subtracted from the observed photospheric magnetogram, and the resultant photospheric magnetogram was used to extrapolate the ambient potential field as the background. Third, they embedded the MFR and the associated magnetic field into the extrapolated background field, and an overall configuration of the magnetic structure including an MFR was then obtained. The advantage of this approach over the others is that the contribution of the longitudinal field within the footpoint area of the MFR to the overall magnetic configuration is just considered once correctly, and that the distribution of the overall magnetic configuration at the photosphere is consistent with the observed photospheric magnetogram. \citet{2019ApJ...884L...1G} confirmed later that this approach indeed works well.

In this work, we employ the RBSL by embedding an MFR into a potential field  to model the 3D magnetic structure of a large-scale IF that is partially located in a decaying and diffuse active region.  Section ~\ref{sec:instruments and Data} describes the instruments and data involved in our work. Section  ~\ref{sec:Observations and Modelling Methods}  lays out the observed properties of the filament, and the method applied to construct the NLFFF magnetic field configuration that includes a filament. Our results and comparisons with observations are presented in Section ~\ref{sec:Modelling Results}. We discuss these results and the related comparisons in Section  ~\ref{sec:Discussions}, and finally summarize this work and draw our conclusions in  Section ~\ref{sec:Conclusions}.

\section{Observational data and instruments}
\label{sec:instruments and Data}

\begin{figure*}
\centering
	\includegraphics[width=1.0\textwidth]{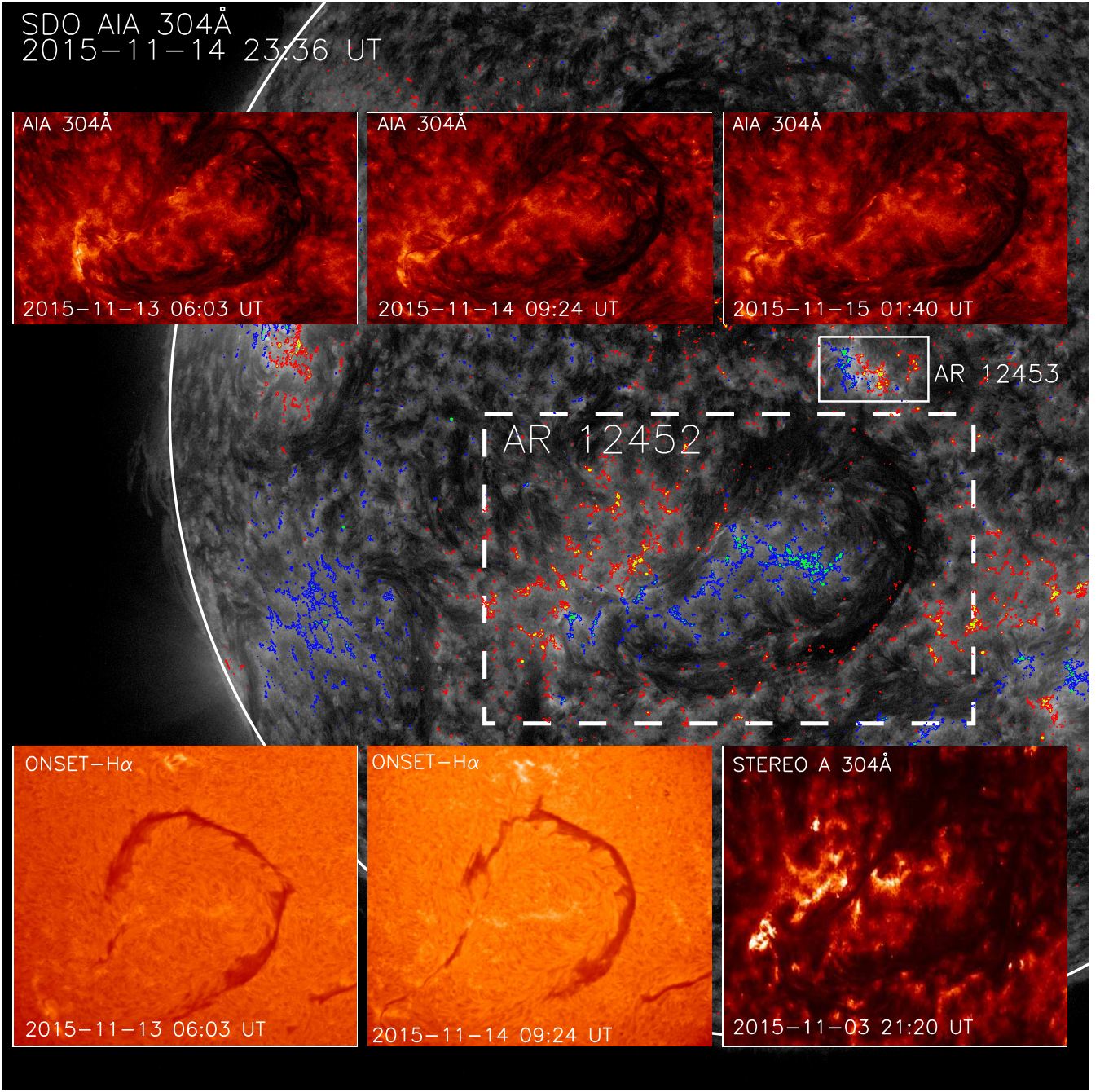}
    \caption{Image collage of the horse-shoe-like filament and its nearby environment observed in different wavelengths at different times by different instruments both in space and on ground. The big image in background is the composite of an AIA 304~\AA \ image and the contours of longitudinal magnetic field of the photosphere in the same region with strengths of $\pm$ 100~G (blue/red) and $\pm$ 500~G (green/yellow). The white arc represents the solar limb. The dashed white lines surround the area where the extrapolation of the coronal mangetic field was performed. The small panels over the background were taken in AIA 304~\AA, STEREO-A 304~\AA, and ONSET H$\alpha$, respectively, over the time interval between 2015 November 3  and  November 15.}
    \label{figure1}  
\end{figure*}

\begin{figure*}
\centering
	\includegraphics[width=0.7\textwidth]{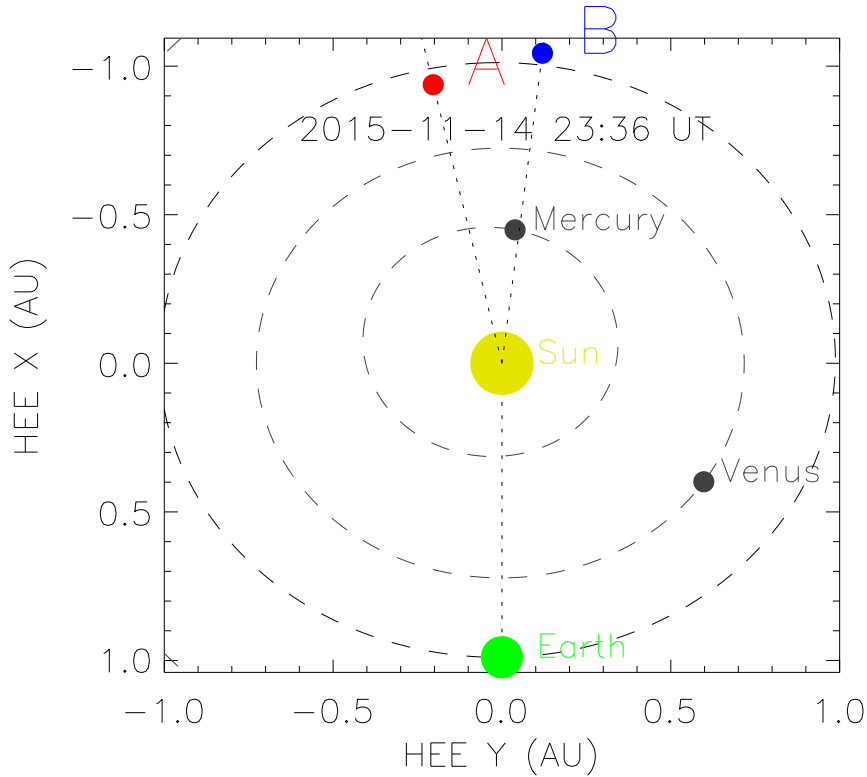}
    \caption{Positions of STEREO-A and -B at 23:36 UT on 2015 November 14 in the HEE coordinate system. }   
    \label{figure2}   
\end{figure*}

A large-scale IF of about  8.5 $\times$ $10^{5}$ km appeared around the active region AR12452 (Figure ~\ref{figure1}). The filament was observed by the Atmospheric Imaging Assembly (AIA) and the Helioseismic and Magnetic Imager (HMI) on board Solar Dynamics Observatory (SDO, \citealt{2012SoPh..275....3P}), the Extreme Ultraviolet Imager (EUVI) in the instrument suite of the Sun Earth Connection Coronal and Heliospheric Investigation (SECCHI) on board Solar TErrestrial RElations Observatory Ahead (STEREO-A, \citealt{2008SSRv..136....5K}), and the ground-based Optical and Near infrared Solar Eruption Tracer (ONSET, \citealt{2013RAA....13.1509F}).

AIA  is an instrument that consists of seven Extreme Ultraviolet (EUV)  and three Ultraviolet (UV) and white-light channels and provides the solar corona's full-disk images with pixel size of $0.6{''}$ and cadence of 12 seconds \citep{2012SoPh..275...17L}, and it was designed to study the evolution of the Sun's dynamic magnetic field and its interaction with surrounding plasma. The EUV full-disk images in seven channels are used in this work. HMI is an instrument which was designed to study the oscillations and the magnetic field on the solar surface and provides the full-disk vector magnetogram with a time cadence of 12 minutes \citep{2012SoPh..275..207S,2012SoPh..275..229S,2014SoPh..289.3483H}. The magnetograms are rescaled down to the AIA plate scale of 0.6 arcsec, and the roll and alignment are adjusted to those of AIA \citep{2012SoPh..275...17L}. The ONSET is a multi-wavelength telescope that  conducts observations in He I 10830  \AA  \ , H$\alpha$  and white-light at 3600 \AA  \  and 4250 \AA  \ in a full-disk observation mode or a partial-disk observation mode. In partial-disk observation mode, ONSET provides partial-disk images with  high spatial resolution ($\sim$ $1{''}$ or better) and  high time cadence (better than 1~s,  \citealt{2013RAA....13.1509F}). The data taken in partial-disk observation mode are used here. SECCHI/EUVI observes the chromosphere and the low corona in four different  wavelengths between 171 \AA  \ and 304 \AA  \ with a field of view (FOV) out to 1.7 solar radii \citep{2004SPIE.5171..111W,2008SSRv..136...67H}.

Observational data were selected based on the following criteria. First, the target region must be located near the disk center in order to minimize the  error in path measurements. Second, the image from AIA, HMI, and ONSET should be taken roughly simultaneously. Third, the impact of the emerging small-scale active region AR12453 (Figure ~\ref{figure1}) on the upper right of the target region  on the  main structure  of the filament should be as small as possible.

On the  basis of these principles, a data set that includes seven AIA images and an ONSET H$\alpha$ image at 09:24 UT on 2015 November 14 was selected for measuring the width of the filament,  another data set that includes an image in AIA 304 \AA  \ and an HMI vector magnetogram at 23:36 UT on 2015 November 14  was selected for modelling the pre-eruptive magnetic configuration and for the related  analyses.

The background image in Figure ~\ref{figure1} displays the composite of the image of the overall magnetic structure around the region observed by SDO/AIA in 304~\AA \ and the contours of the LOS magnetic field observed by SDO/HMI. The color curves in the figure describe the contours of magnetic fields of $\pm 100$~G (blue/red) and of those of $\pm 500$~G  (green/yellow). Inset panels near the upper and the lower edges in Figure ~\ref{figure1} display the filtergrams of a local region surrounded by the dashed box in the big image observed in ONSET/H$\alpha$, AIA/304\AA, and SECCHI/304~\AA, respectively, at different times.

The filament appeared over the east limb of the Sun in the AIA  FOV for the first time at 02:00 UT on 2015 November 7. It took 8 days for the filament to move from east limb to the central meridian on the solar disk as a result of the solar rotation, and the filament started eruption at about 21:00 UT on 2015 November 15. Observations of SECCHI/EUVI of STEREO-A in 304~\AA \   (Figure ~\ref{figure1}) show that the filament actually existed already on 2015 November 3 four days before it appeared in the AIA FOV. This suggests that the filament existed stably for a dozen days before erupting. The position of STEREO-A in the heliocentric earth ecliptic (HEE) coordinate system is shown in Figure ~\ref{figure2}. In addition, small panels in Figure ~\ref{figure1} also show that the filament observed in H$\alpha$ evolved during the period in which it passed through the solar disk, but the overall shape of the filament observed in 304~\AA \  showed almost no change. This indicates that the global magnetic structure of the filament channel is stable and in the process of a quasi-static evolution. Consequently, a static NLFFF model to construct the magnetic structure of the filament before its eruption is appropriate here.

From the contours in Figure ~\ref{figure1}, we realize that the magnetic field in AR 12452 is weak with very few confined area possessing  a magnetic field of strength up to 500 G. This indicates that AR 12452 was a decaying and diffuse active region. Therefore, magnetic structures of the filament in this active region cannot be constructed by using the traditional NLFFF extrapolation method \citep{2004SoPh..219...87W}. As formulated by \citet{2018ApJ...852L..21T}, the RBSL method generalizes the approach of \citet{2014ApJ...790..163T} for MFRs of arbitrary shapes and so, by construction, includes the MFR embedding. Thus we here employ the method of \citet{2018ApJ...852L..21T} to construct the magnetic structure of the filament in AR 12452. The RBSL code we adopt here was implemented by \citet{2019ApJ...884L...1G}.

\section{Approaches to reconstructing the magnetic structure of the filament}
\label{sec:Observations and Modelling Methods}

\begin{figure*}
\centering
	\includegraphics[width=1.0\textwidth]{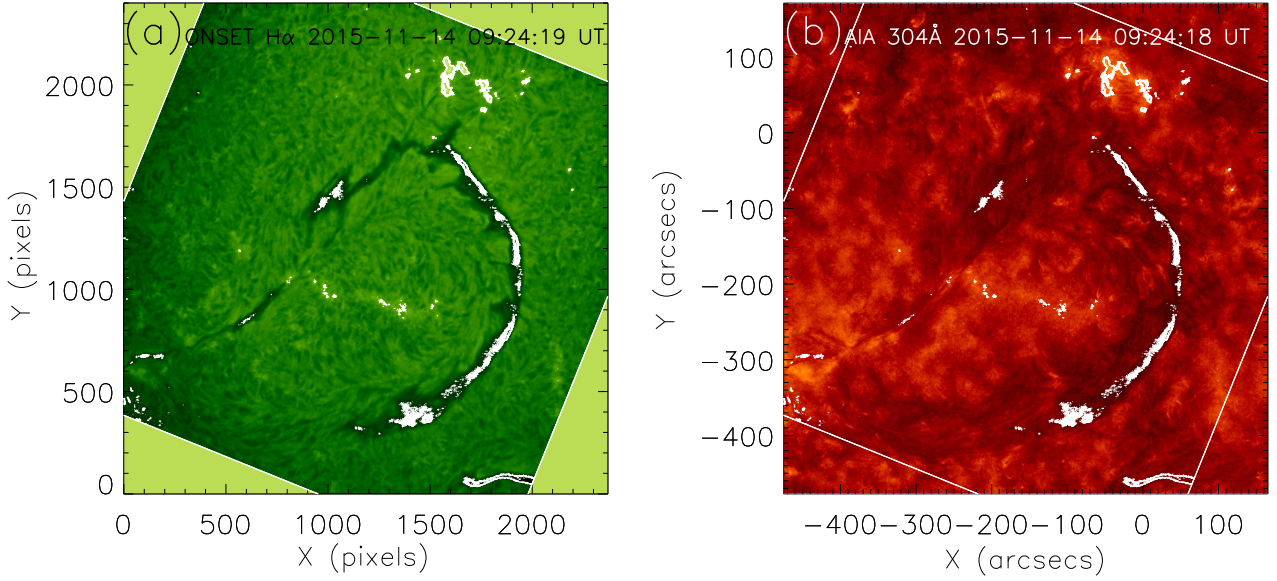}
    \caption{Co-aligned ONSET H$\alpha$ image (a) and AIA 304 \AA \ image (b). The white patches in one image denote the features that could be recognized in another image.}  
    \label{fig:figure3}
    
\end{figure*}

\begin{figure*}
\centering
	\includegraphics[width=1.0\textwidth]{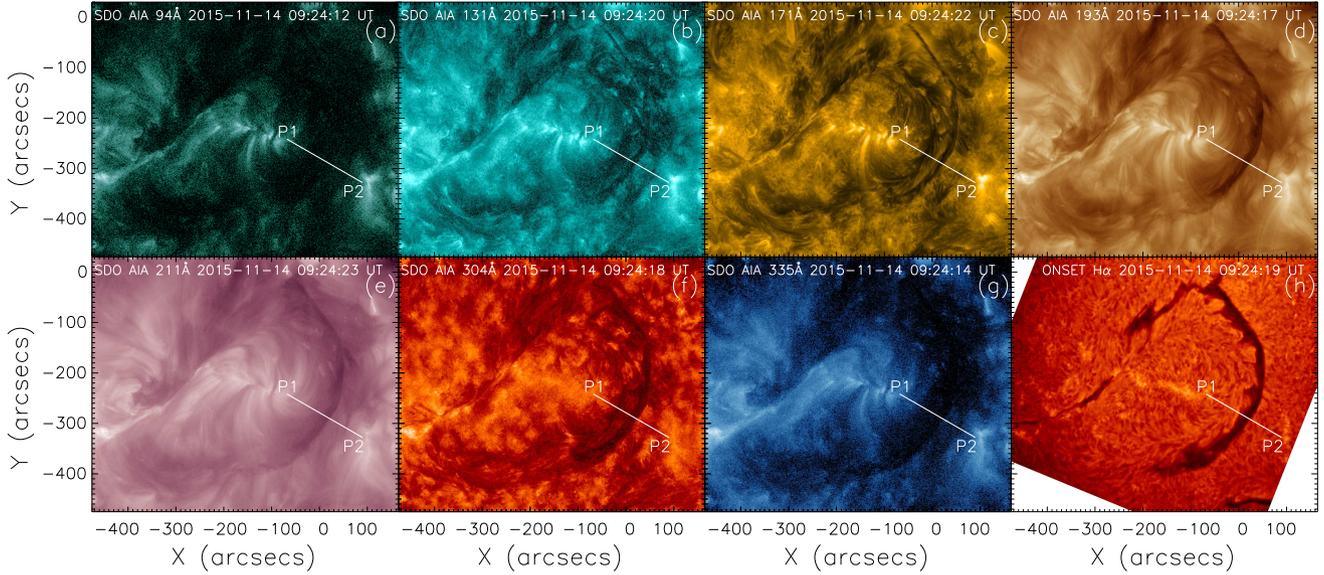}
    \caption{Collage of co-aligned ONSET H$\alpha$ image and AIA images in different wavelengths of 94~\AA (a), 131~\AA (b), 171~\AA (c), 193~\AA (d), 211~\AA (e), 304~\AA (f), 335~\AA (g), ONSET H$\alpha$ (h). The white line, P1P2, in each panel marks the route along which the brightness distribution of the image is deduced for Figure ~\ref{figure5}.}  
    \label{figure4}
    
\end{figure*}

We use a static NLFFF model to construct the magnetic structure of the filament before its eruption.
The ONSET H$\alpha$ images  show that the filament has left-bearing barbs extending from  its main body. In addition, it can also be concluded that the filament is sinistral as the axial component of the magnetic field points to the left when it is seen by an observer standing on the positive end of the filament.

\begin{figure*}
\centering
	\includegraphics[width=1.4\columnwidth]{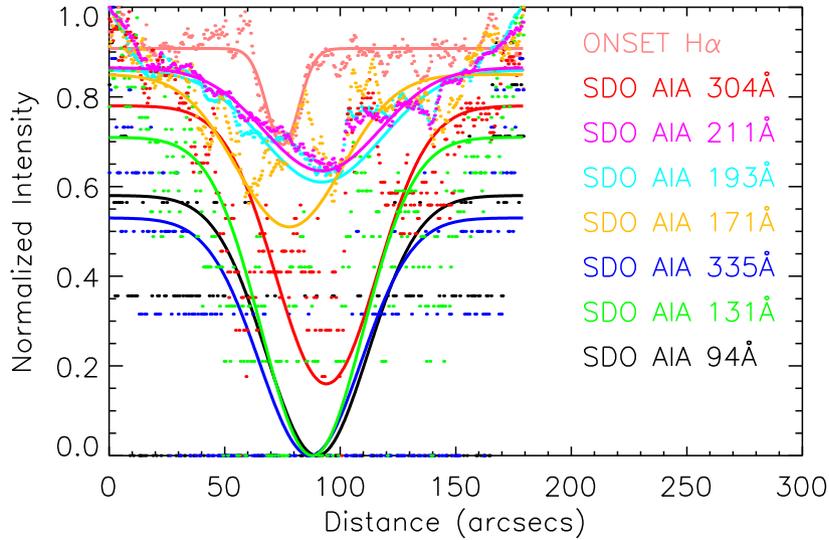}
    \caption{Distributions of the brightness in various wavelengths (dots) and the corresponding Gaussian fits (continuous curves) along the white line P1P2 in Figure ~\ref{figure4}. All the values of the brightness have been normalized.}
    \label{figure5}
    
\end{figure*}

\begin{figure*}
\centering
	\includegraphics[width=0.75\textwidth]{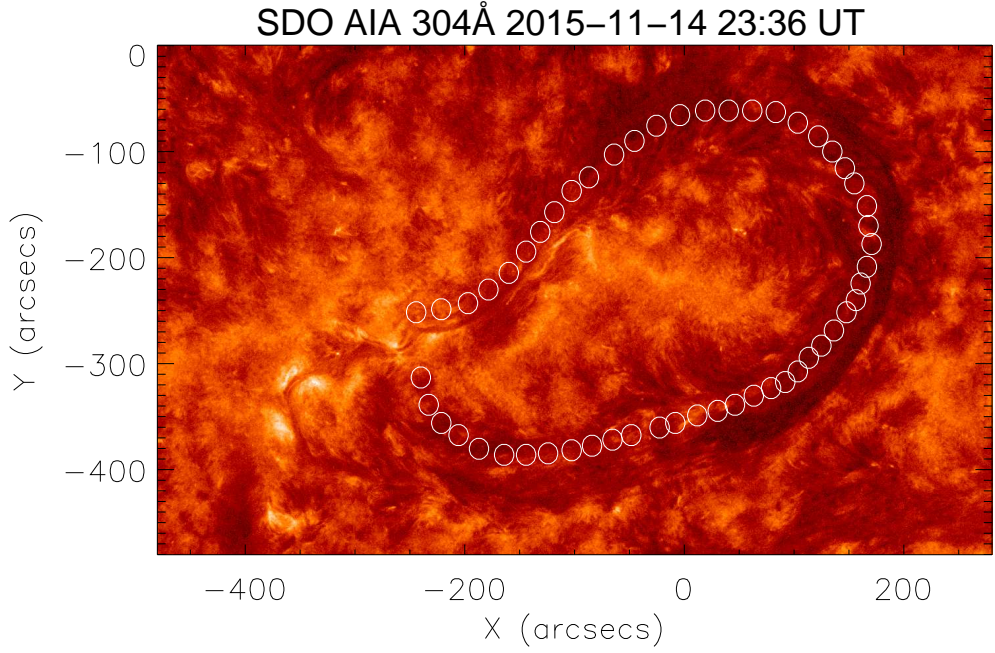}
    \caption{The 304  \AA \ image observed by SDO/AIA  at 23:36 UT on 2015 November 14 with the filament axis marked by a set of open  circles.}
    \label{figure6}
    
\end{figure*}

\begin{figure*}
\centering
	\includegraphics[width=0.75\textwidth]{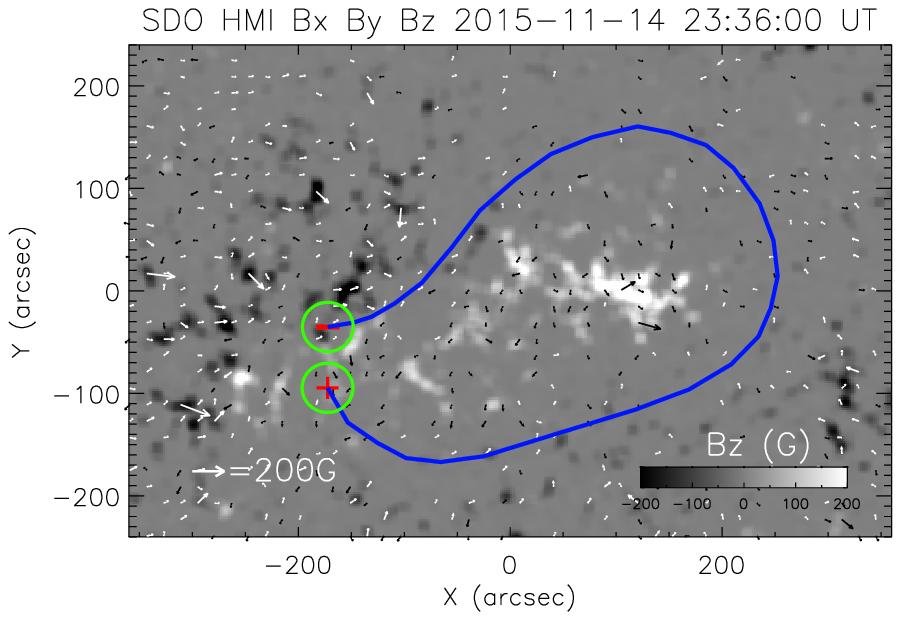}
    \caption{SDO/HMI vector magnetic field obtained at 23:36 UT on 2015 November 14 overlaid with the path of the filament axis (blue curve), the red plus and minus signs indicate the two footprints of the filament axis. The green circles mark the periphery of the filament footprints.}
    \label{figure7}   
\end{figure*}

According to the existing observational results, \citet{1998SoPh..182..107M} concluded that the sinistral/dextral filament always had left/right bearing barbs, and this law seemed true for a couple of decades until \citet{2010ApJ...714..343G}  found that a dextral filament can also feature left-bearing barbs. \citet{chen2014ApJ...784...50C} further confirmed the findings of \citet{2010ApJ...714..343G}, namely  a filament that follows  Martin's law has a MFR  configuration, while a filment that fails to follow  Martin's law has a sheared arcade magnetic configuration. \citet{2017ApJ...835...94O} performed a statistical study to address this issue, and they  found that 60$\%$ of AF are supported by flux rope, 40$\%$ are supported by sheared arcade. For IF, these two values are 91$\%$ and 9$\%$, respectively, while 96$\%$ and 4$\%$ for QF.

The filament studied here is sinistral and has left-bearing barbs, so it should possess an  MFR  configuration according to  \citet{2010ApJ...714..343G} and \citet{chen2014ApJ...784...50C}, and thus we could apply  the MFR embedding method  to construct the  magnetic structure of the filament.

The MFR in our model is calculated by using the RBSL, which is given by   \citet{2018ApJ...852L..21T} as

\begin{equation}
    \boldsymbol{B}_\textbf{MFR}=\nabla\times\boldsymbol{A}_I+\nabla\times\boldsymbol{A}_F,
	\label{eq:quadratic}
\end{equation}

\begin{equation}
    \boldsymbol{A}_I(\boldsymbol{x})=\frac{\mu{I}}{4\pi}\int_{\textit{C}\cup{\textit{C}}^{*}}K_I(\boldsymbol{r})\boldsymbol{R}^\prime(l)\frac{dl}{a(l)},
	\label{eq:quadratic}
\end{equation}

\begin{equation}
    \boldsymbol{A}_F(\boldsymbol{x})=\frac{F}{4\pi}\int_{\textit{C}\cup{\textit{C}}^{*}}K_F(\boldsymbol{r})\boldsymbol{R}^\prime(l)\times\boldsymbol{r}\frac{dl}{a(l)^2},
	\label{eq:quadratic}
\end{equation}
where $\boldsymbol{R}^\prime(l)$ = $\textit{d}\boldsymbol{R}$/$\textit{d}l$, $\boldsymbol{r}=(\boldsymbol{x}-\boldsymbol{R}(l))/a(l) $, while $\textit{C}$, $\textit{l}$, $\boldsymbol{R}(l)$ and $\textit{a}(\textit{l})$ are axis path, arc length, radius-vector and cross-section radius of a thin MFR, respectively. $\textit{C}^{*}$ is the subphotospheric counterpart of $\textit{C}$. The field $\boldsymbol{A}_I(\boldsymbol{x})$ and its curl represent the axial vector potential and the azimuthal magnetic field, respectively, which are generated by a net current \textit{I}. The field $\boldsymbol{A}_F(\boldsymbol{x})$ and its curl represent the azimuthal vector potential and the axial magnetic field, respectively, which are generated by a net flux \textit{F}. Details of kernels $K_I(\boldsymbol{r})$ and $K_F(\boldsymbol{r})$ of intergrals were given by  \citet{2018ApJ...852L..21T}. Four free parameters are involved in RBSL, which are the minor radius of the flux rope, $\textit{a}$, the path, $\textit{C}$, the net magnetic flux, \textit{F}, and the net electric current, \textit{I}. Hence, we first need to determine these parameters in order to construct the MFR configuration of the filament with RBSL.

In our model, the value of $a$ is determined  by the half-width of the filament, which could be deduced from images  of AIA and H$\alpha$ images. However, the images taken by AIA and ONSET are not aligned to one another, and we need to perform alignment before further analyses, for which the technique developed by \citet{2022A&A...659A..76W}  is utilized.  Figure ~\ref{fig:figure3} shows the composite of an ONSET H$\alpha$ image and an AIA 304 \AA \ image that were taken almost simultaneously. To reveal the relative locations of the filament seen in different wavelengths, we display both original images of H$\alpha$ and 304 \AA \ in Figures ~\ref{fig:figure3}(a) and ~\ref{fig:figure3}(b), respectively, with those white patches outlining the features seen in another wavelength. We notice that the global features of the filament observed in different wavelengths are generally colocated with one another in space, which implies also that the alignment conducted here is basically successful.

Based on this result, we are able to measure the width of the filament. The measurement is performed along the white line as shown in Figure  ~\ref{figure4} at a position where the filament looks uniform, avoiding the location where either the spine has barbs or the filament looks too thin. Figure  ~\ref{figure5} plots distributions of the normalized brightness of the filament along the white line in various wavelengths, as well as the associated Gauss-fitting-curves. The full widths at half maximum (FWHM) of these  curves  give the corresponding widths of the filament observed in given wavelengths. Taking the half of the average of all the FWHMs gives the minor radius, $a=0.026$~R$_{\odot}$, of the filament, which is one of the free parameters used in RBSL.

The path of MFR in our model is also constrained by  observations of the filament. We use the path of the filament axis as the path of MFR, which should be a 3D curve. Because STEREO-B has lost communications since October 2014, the 3D path of the filament cannot be constructed on the basis of the data from STEREO-A and -B via the approach of the triangulation developed by \citet{2009Icar..200..351T}. Meanwhile, STEREO-A was almost at the opposite side of the Sun relative to SDO (Figure ~\ref{figure2}), and they cannot observe the filament simultaneously. So it is not possible to construct the 3D path according to observations of STEREO-A and SDO, either. Therefore, we have to resort to other ways to determine the 3D path of the filament, which is described below.

As shown in Figure ~\ref{figure6}, we first determine the coordinates $(X, Y)$ in arcsec of the filament path on the AIA 304 \AA \ image taken at 23:36 UT on 2015 November 14 when the filament is close to the solar disk center. Then we use  routine "xy2lonlat.pro" in the Solar SoftWare (SSW) to convert the  coordinates on the solar disk  into the heliographic coordinates. At the same time, we select some features on the filament observed in AIA 304 \AA \ when it just entered the FOV of AIA on the east limb of the Sun, and measure their height away from the solar surface, which is regarded as the $Z$ component of 3D path. With the coordinates $(X, Y)$ of all these features (see those circles in Figure ~\ref{figure6}) obtained, the 3D coordinates  for describing the path $\textit{C}$ along the filament axis required in RBSL are collected.

With $a$ and $C$ being determined, the third free parameter, $F$, can be obtained consequently. First of all, after the projection effects have been corrected, the vector magnetogram in the region of interest is displayed in Figure ~\ref{figure7} with the blue curve tracking the axis of the filament. The red symbols, "+" and "$-$" in Figure ~\ref{figure7}  specify the location of the filament footpoints on the photosphere. The location of the reference point that is the center of the magnetogram in Figure ~\ref{figure7}  is located at E5$^{\circ}$.4 and S10$^{\circ}$.8. The  magnetogram can cause significant errors when it is directly used for extrapolating a force-free magnetic field because the  photospheric magnetic field is far from being force-free \citep{1995ApJ...439..474M,2008SoPh..247..269M}. \citet{2006SoPh..233..215W} found that appropriate preprocessing can greatly reduce these errors. Thus, the projection-corrected vector magnetic field is also preprocessed to remove the Lorentz force and torque \citep{2006SoPh..233..215W,2008JGRA..113.3S02W}.

Second, the heliospheric coordinates of the points on the filament axis are converted into the Cartesian coordinates with the origin located at the reference point. The axes, \textit{x}, \textit{y}, \textit{z}, in the local Cartesian coordinate system are  westward, northward and radial, respectively, and plane $z=0$ is tangent to the solar surface at the reference point. The path in the local Cartesian coordinate system is illustrated by the blue solid line in Figure  ~\ref{figure7}.

Third, with each end of the path as the center, we draw two circles  with $\textit{a}$  as the radius. The two green circles plotted in Figure  ~\ref{figure7} present the regions to which the footprints of the filament anchor. Finally, we measure the longitudinal magnetic flux within the two circles and find that $F_+$ = 1.27 $\times$ $10^{20}$ Mx and $F_-$ = $-$2.36 $\times$ $10^{20}$ Mx. The magnetic flux of the MFR here is thus taken as \textit{F} = (|$F_+$| + |$F_-$|) /2 = 1.82 $\times$ $10^{20}$ Mx.

With the value of \textit{F} obtained, the fourth physical parameter for MFR, \textit{I}, is evaluated via (see also \citealt{2018ApJ...852L..21T}) \textit{F} = $\pm$3$\mu_0$\textit{I}\textit{a} / (5$\sqrt{2}$), where $\mu_0$ is the magnetic permeability of vacuum, and \textit{I} could be positive or negative depending on the sign of the magnetic helicity. In the present case, we already know that the  chirality of the filament is sinistral, which indicates that the magnetic field possesses a positive magnetic helicity. Hence, the sign of the parameter \textit{I} is taken as positive in our model, and the value of \textit{I} is calculated by \textit{I} = (5$\sqrt{2}$\textit{F}) / (3$\mu_0$\textit{a}) = 
2.45 $\times$ $10^{11}$ A. So far, four parameters have been obtained for constructing the magnetic configuration including an MFR via RBSL approach, and detailed descriptions of how such a configuration is built up are then given below.

\begin{figure}
\centering
	\includegraphics[width=\columnwidth]{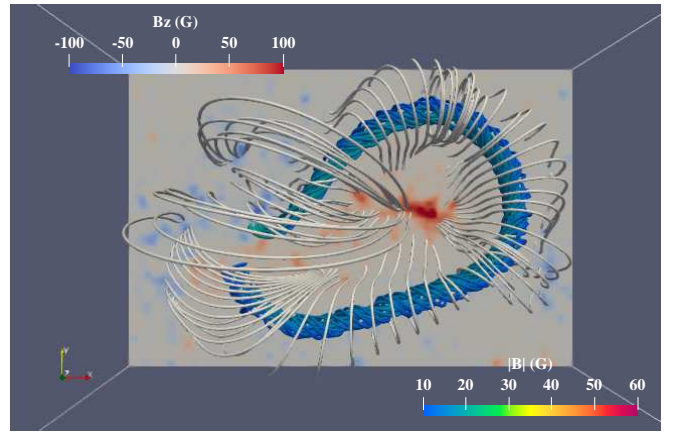}
    \caption{Magnetic configuration including the MFR constructed via MFR embedding and the potential background field extrapolated via the Green's function method according to the data obtained at 23:36~UT on November 15, 2015. This is the topology obtained before a final relaxation step to a more force-free setup.}
    \label{figure8}
    
\end{figure}

First, a mirror path of  $\textit{C}$ is used to close  the electric current circuit in the whole space as required by RBSL  \citep{2018ApJ...852L..21T}; then we deduce the magnetic field of the MFR determined by the above four free parameters according to the RBSL.
Second, a background magnetic field of the target region is deduced by the potential field (PF) extrapolation based on the Green's function method \citep{1969SoPh....6..442S,1982SoPh...76..301S}. To keep the longitudinal component of the magnetic field on the photosphere unchanged after the construction is performed, the corresponding longitudinal component of the MFR magnetic field within the two circles, $\textit{B}_{zMFR}$, shown in Figure ~\ref{figure7} needs to be subtracted. Then use $\textit{B}_z$$-$$\textit{B}_{zMFR}$ on the photosphere as the boundary condition to extrapolate  the background PF. Here $\textit{B}_z$ is given by the observed magnetogram, and the extrapolation is conducted by using the Message Passing Interface Adaptive Mesh Refinement Versatile Advection Code (MPI-AMRVAC; \citealt{2012JCoPh.231..718K}; \citealt{2014ApJS..214....4P}; \citealt{2018ApJS..234...30X}; \citealt{KEPPENS2021316}).
Then, a non-potential magnetic configuration, as shown in Figure ~\ref{figure8}, is obtained by linearly combining the MFR and the PF magnetic fields. The resultant configuration is further relaxed to a more realistic NLFFF state via a magnetofriction approach (e,g, see also \citealt{2016ApJ...828...82G,2016ApJ...828...83G}) exploiting MPI-AMRVAC with the configuration in Figure ~\ref{figure8} as the initial condition and the vector magnetic field shown in  Figure ~\ref{figure7} as the boundary condition at the bottom.

\section{Results}
\label{sec:Modelling Results}

\begin{figure}
\centering
	\includegraphics[width=\columnwidth]{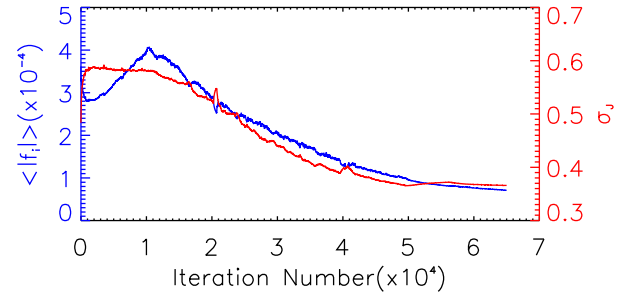}
    \caption{Variations of values of $<|f_{i}|>$ (blue) and $\sigma_J$ (red) versus the iteration number in the process of relaxation via magnetofriction.}
    \label{figure9}
    
\end{figure}

\begin{figure*}
\centering
	\includegraphics[width=1.0\textwidth]{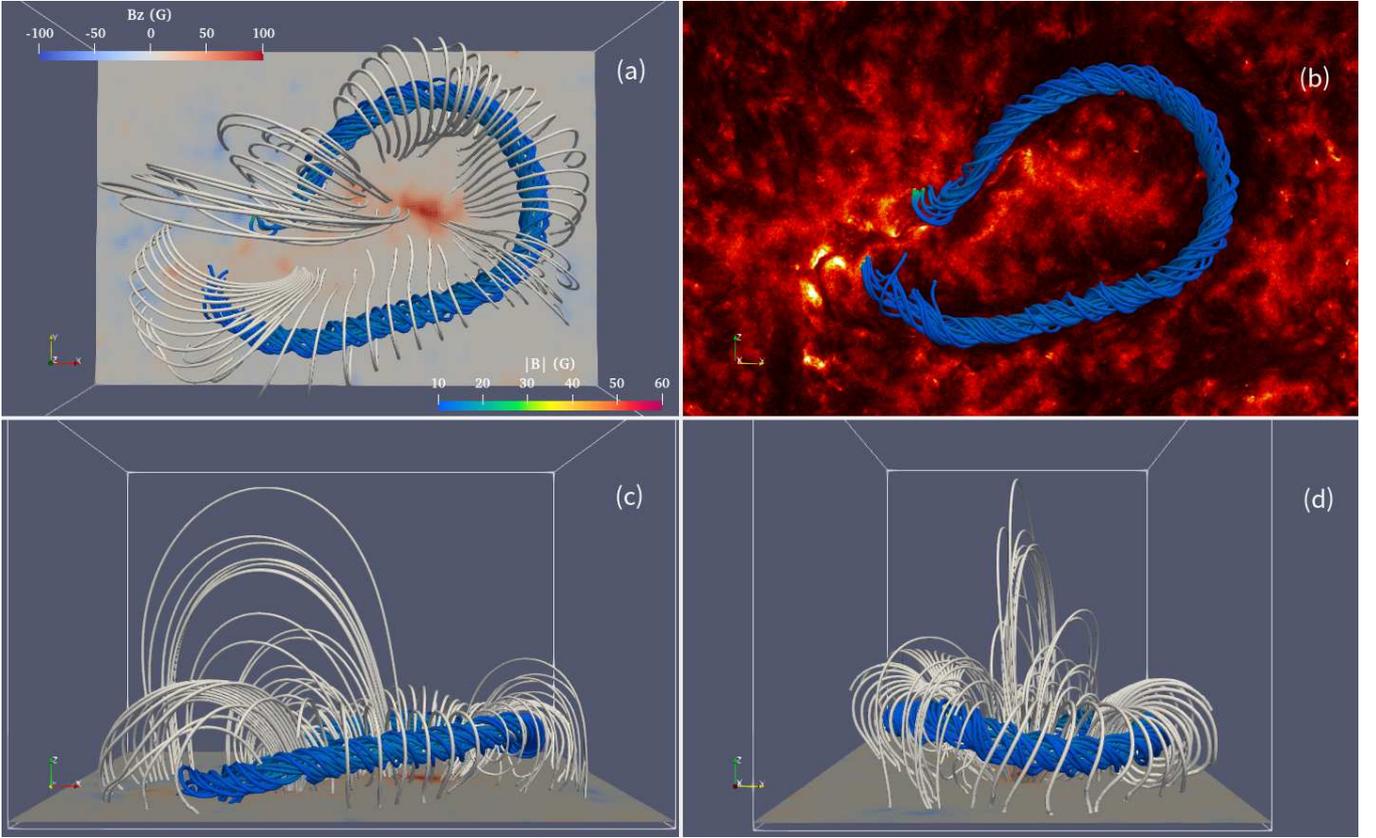}
    \caption{(a) The relaxed magnetic field after a magnetofriction process associated with 65,000 iterations. The axes, \textit{x}, \textit{y} and \textit{z}, are  westward, northward and radial, respectively. Blue curves are magnetic field lines tangling inside the MFR, and white curves are the field lines outside the MFR. The LOS is in the direction opposite to the \textit{z}-axis.  (b) The same MFR superposed on an AIA 304 \AA \ image that was observed  at 23:36 UT on 2015 November 14. (c) and (d) Same as (a) but viewing in $y$- and $-x$-directions, respectively. An associated animation  is available. }  
    \label{figure10}
    
\end{figure*}

\begin{figure*}
\centering
	\includegraphics[width=1.0\textwidth]{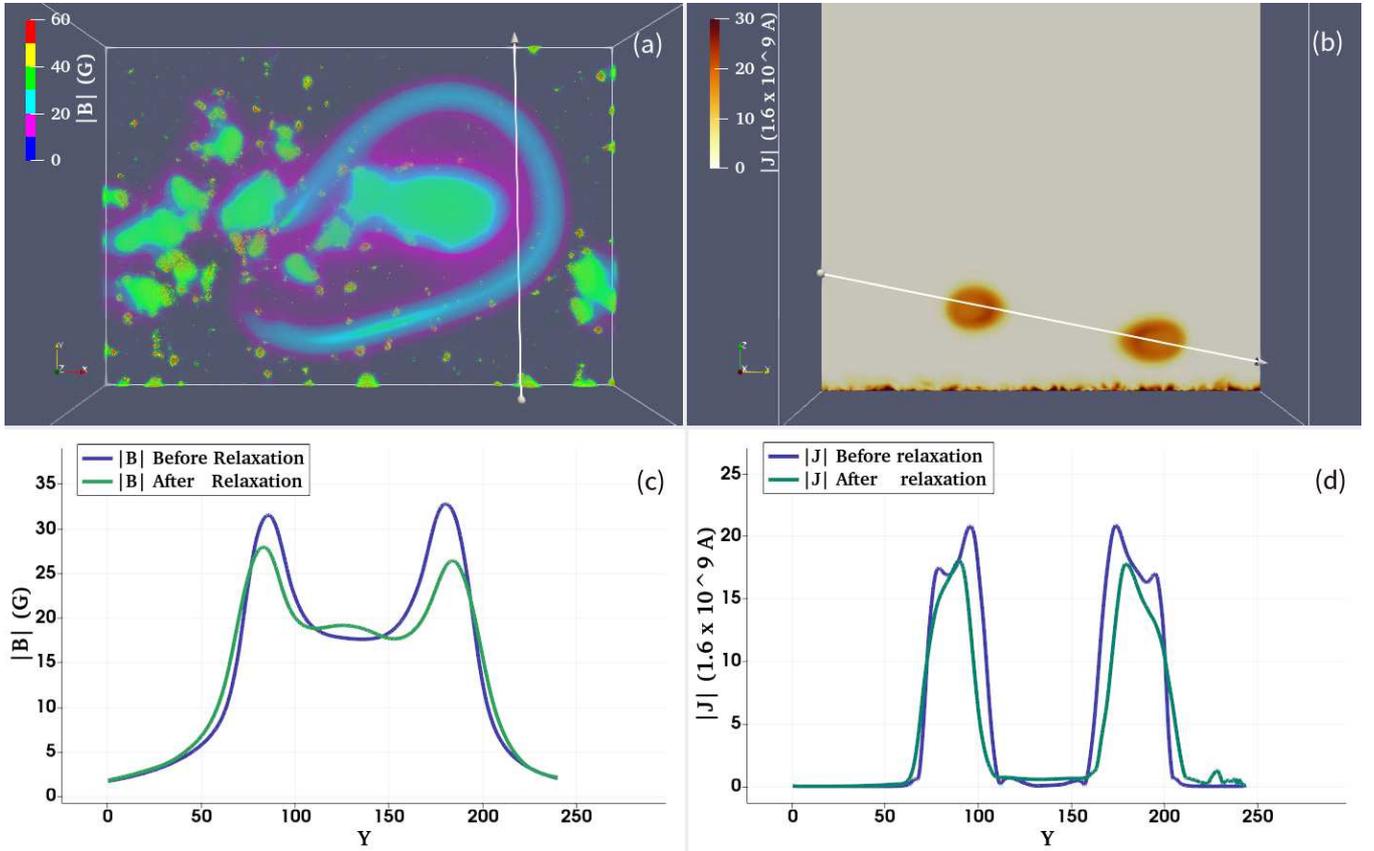}
    \caption{(a) 3D strength contours of the magnetic field in the region surrounding the MFR shown in Figure ~\ref{figure10}(a) with colors denoting different strengths of magnetic field. The white line with end points at $(-32^{\prime\prime}.5, -480^{\prime\prime}.0, 95^{\prime\prime}.8)$ and $(-32^{\prime\prime}.5, 0^{\prime\prime}.0, 55^{\prime\prime}.3)$ is used to mark a plane on which distributions of the magnetic field and the electric current could be displayed. An associated animation is available. (b) The distribution of the electric currents on the plane denoted in panel (a). The white line goes through the cross-sections of the MFR and is used to study distributions of the magnetic fields (c) and the electric currents (d) inside the MFR before (blue curves) and after (green curves) the relaxation, respectively.}
    \label{figure11}
    
\end{figure*}

\begin{figure*}
\centering
	\includegraphics[width=1.0\textwidth]{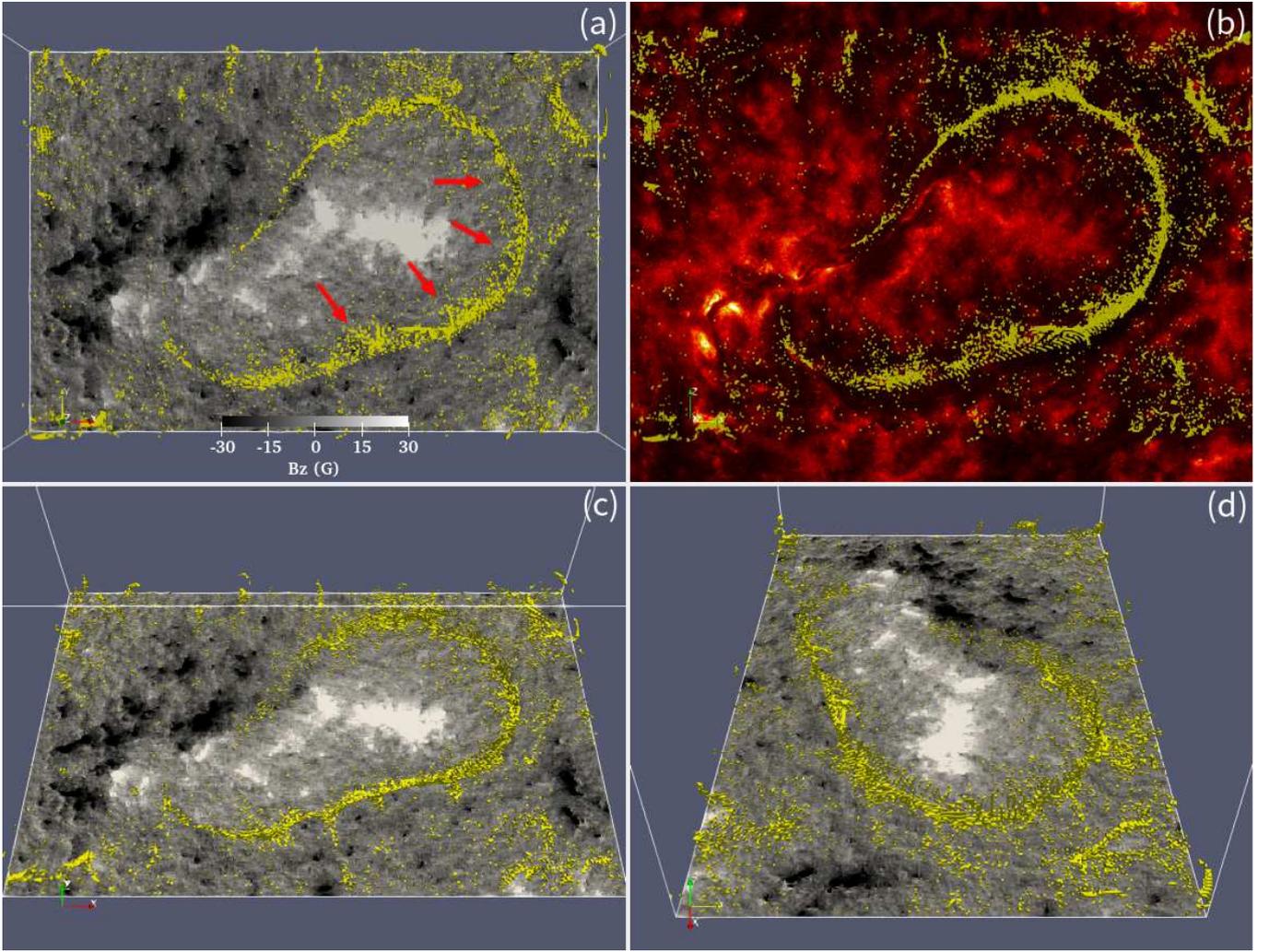}
    \caption{Panels (a) and (b) are the same as Figures  ~\ref{figure10}(a) and  ~\ref{figure10}(b), but for the distributions of the magnetic dips in the same area. (c) Viewing the same area as that in panel (a) along an LOS in the $zx$-plane, and the angle between LOS and the photosphere is 65$^{\circ}$. (d) The same as panel (c) but the LOS is in the $yz$-plane, and the angle between the LOS and the photosphere is 40$^{\circ}$. An associated animation is available.}
    \label{figure12}
    
\end{figure*}

\begin{figure*}
\centering
	\includegraphics[width=1.0\textwidth]{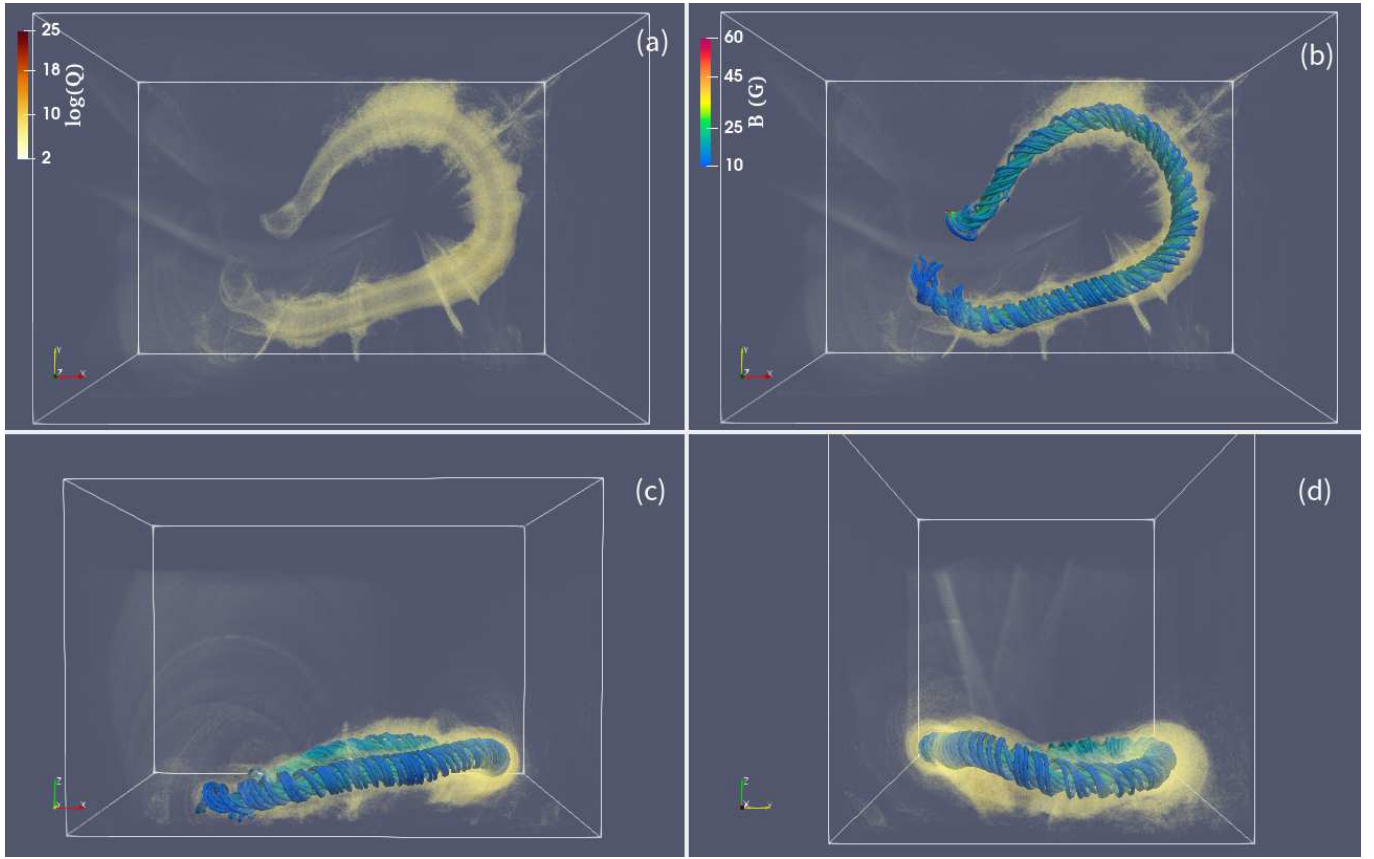}
    \caption{(a) Distribution of the QSL in the same region as that in Figure ~\ref{figure10}(a). Panels (b)-(d) display the MFR too, as well as the viewpoints in (c) and (d) are the same as those in Figures ~\ref{figure10}(c) and ~\ref{figure10}(d). An associated animation  is available.} 

    \label{figure13}
    
\end{figure*}

\begin{figure*}
\centering
	\includegraphics[width=1.0\textwidth]{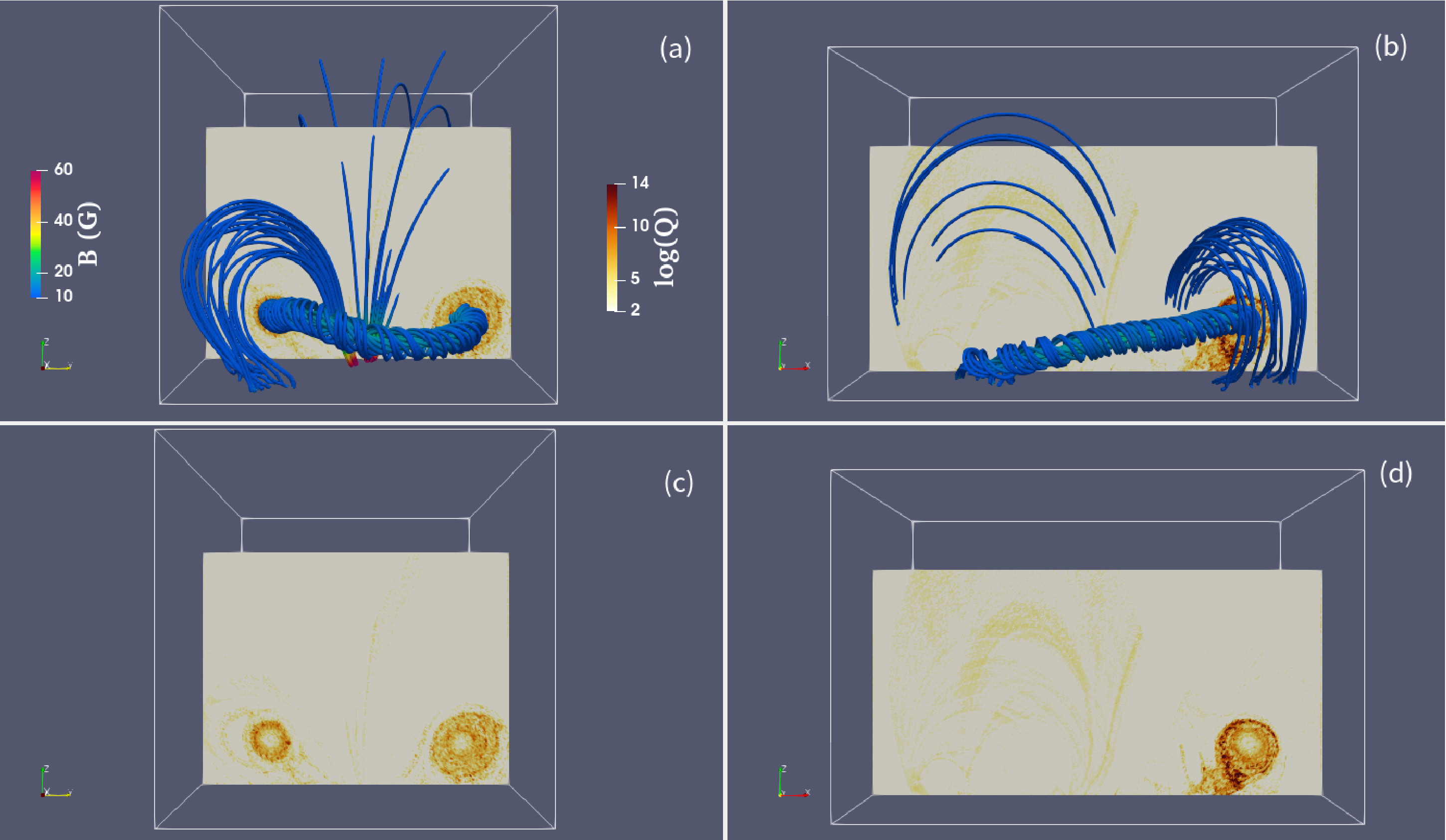}
    \caption{(a) Global configuration including the MFR and the associated background magnetic field (blue curves), as well as contours of the QSL on plane $X=-154^{\prime\prime}.1$. (b) Same as panel (a), but for contours of the QSL on plane $Y=-292^{\prime\prime}$. Panels (c) and (d) display the same QSL contours as in (a) and (b) without MFR and the background field in order to reveal detailed features of QSL.}
    \label{figure14}
    
\end{figure*}

To assess the imposed divergence-free and the force-free constraints  of the constructed magnetic field in the magnetofrictional relaxation, we evaluate two parameters,  $<|f_i|>$  and $\sigma_J$, and check how close they are to zero. In the present work, since the size of the grid is uniform,  $<|f_i|>$ is the average of the absolute value of the fractional magnetic flux change over all the 6 sides of every cubic-grid in 3D calculations, and $\sigma_J$ is the current-weighted average of the sine of the angle between the current and the magnetic field,  respectively. Changes in  $<|f_i|>$ and  $\sigma_J$ with respect to the number of iterations are presented in  Figure ~\ref{figure9}, in which the blue curve is  for  $<|f_i|>$ and the red curve for  $\sigma_J$. It can be  seen from Figure ~\ref{figure9} that both $<|f_i|>$ and  $\sigma_J$ decrease eventually in the process of relaxation represented by the increasing iteration number, and then they asymptotically approach to a finite value, respectively, such that  $<|f_i|>$ = 7.1 $\times$ $10^{-5}$ and $\sigma_J$ = 0.36. This indicates that the divergence-free condition of the constructed magnetic field reaches an acceptable level, and that the force-free condition has largely improved after the relaxation procedure. The value of $\sigma_J$ = 0.36  is already very close to that in the best case acquired by \citep{2016ApJ...828...82G} although smaller $\sigma_{J}$ could still be expected.

Figure ~\ref{figure10} displays the constructed MFR magnetic configuration after relaxation seen from the top with some magnetic field lines around (Figure ~\ref{figure10}(a)), a composite of MFR and the background seen in AIA~304 \AA \ (Figure ~\ref{figure10}(b)), and the configuration seen in $y$- and -$x$-directions, respectively. The difference between the configuration before relaxation (Figure ~\ref{figure8}) and that after relaxation (Figure ~\ref{figure10}(a)) is not apparent, and is not easy to recognize at the first sight. This can be considered as a significant advantage of the MFR embedding approach over the MFR inserting method.

We can see  that the MFR has a right-helical twist. Comparing with the initial magnetic configuration (see  Figure ~\ref{figure8}), we realize that the northern footpoint of the MFR moves to the north, while the southern footpoint of the MFR moves to the south after the relaxation. The displacement of the southern footpoint is larger than that of  the northern one, which is due to the fact that the photospheric field at the northern footpoint region of the filament are stronger than that at the southern one. It should be stressed that this motion is not a real evolution scenario, but the rearrangements in field strength and current which are ensuring that a more force-free field configuration is established. In addtion, the height of the MFR axis varies with position along the filament (see Figure ~\ref{figure10}(c)),  in the middle part of the MFR the height of the MFR axis does not change too much and almost stays  horizontally (see Figure ~\ref{figure10}(d)), and the highest point of the MFR is about   $7.85 \times 10^{4}$ km above the photosphere and the average width of the flux rope is about $3.98 \times 10^{4}$ km,  which is slightly larger than the measured FWHM of the filament.

To produce an image of the magnetic structure/configuration seen by SDO, a back-projection of the above structure  created in the local Cartesian coordinate system needs to be done according to the view-point of SDO via
 the rotation matrix $\mathcal{R}_x$($-$$L_0$)$\mathcal{R}_y$($-$$L$)$\mathcal{R}_x$($L_1$) \citep{2017ScChD..60.1408G}, where $\mathcal{R}_x$ is the elementary rotation matrix with respect to the \textit{x}-axis, $\mathcal{R}_y$ is the elementary rotation matrix with respect to the \textit{y}-axis, $L_0$  is the latitude of the disk center in the image obtained by a specific instrument, $L$ and $L_1$ are the longitude and latitude of the reference point in the vector magnetogram, respectively.

In the present case, the coordinates of ($L_0$, $L$, $L_1$) for  SDO images are $(2.9^{\circ}, -5.4^{\circ}, -10.8^{\circ}$). Figure ~\ref{figure10}(b) displays the composite of an AIA~304 \AA \ image and the MFR that would be seen by SDO at 23:36~UT on 2015 November 14. It can be  seen clearly that the overall morphology of the magnetic structure associated with the constructed MFR is basically consistent with that of the observed filament. The horse-shoe-like  spine is well covered by the field lines of the MFR, which suggests that the construction of the magnetic configuration associated with the horse-shoe-like filament is successful although Figure ~\ref{figure10}(b) indicates a tiny deviation in the north part of the target region.

In Figure ~\ref{figure11}(a), we show the strength of the magnetic field through  different colors with different transparency and notice that, across the MFR, three  regions of different magnetic field strengths exist. The first region is the outside of the MFR, outlined as a pink  horse-shoe-like tube, where the field strength  lies in the range from 7 to 18 G and the lateral spatial variation is the most apparent. The second region is the internal region  of the MFR, shaping as a light blue horse-shoe-like region, in which variation of the magnetic field in space is smoother than that in the first region, and the field is stronger with the strength between 19 and 29 G. The third region is not a single continuous one, but several  separate pieces (see those green areas in Figure ~\ref{figure11}(a)), in which the strength of magnetic field varies from 30 to 34 G.

The strongest magnetic field inside MFR is not located right at the axis of the MFR, but somewhere toward the inner edge of the MFR.  This indicates an asymmetric distribution of magnetic field inside MFR with respect to the axis. We also notice a lateral expansion of the MFR during the relaxation with the average  internal magnetic field weakening from 26 to 23 G.

The MFR evolution in the relaxation process could also be exhibited by the change in the distribution of the electric current inside the MFR. Figure ~\ref{figure11}(b) shows the current distribution in a plane around the MFR before the  relaxation. The plane is located at $X$  = $-32.5^{\prime\prime}$ in the solar  coordinates $(X, Y)$ , which intersects the MFR around the middle part of the MFR, roughly perpendicular to the major axis of the MFR (see the white line in Figure ~\ref{figure11}(a)). It can be seen from Figure ~\ref{figure11}(b) that the current  near the main axis of the MFR is weaker than that  at the edge of the MFR, i.e., the current within the MFR  exhibits a sort of   \textit{hollow} distribution. Similar feature of the current distribution inside the MFR was also reported by  \citet{2008ApJ...672.1209B}.

Evolution of the MFR in the relaxation process could also be represented by the changes in the magnetic field and the electric current on a plane, $X=-32.5^{\prime\prime}$ (see Figure ~\ref{figure11}(b)), during the relaxation. The distributions of the magnetic field strength  and the current intensity along the white line shown in Figure ~\ref{figure11}(b) are given in Figures ~\ref{figure11}(c) and ~\ref{figure11}(d), respectively, in which the blue curve is for the case before the relaxation, and the green one for that after relaxation. Two ends of the white line in Figure ~\ref{figure11}(b) are located at $(-32.5^{\prime\prime}, -480.0^{\prime\prime}, 95.8^{\prime\prime})$ and $(-32.5^{\prime\prime}, 0.0^{\prime\prime}, 55.3^{\prime\prime})$, respectively. Here, the first two values in parentheses correspond to the solar coordinates $(X, Y)$, and the third one is the height from the photospheric surface.

Figure ~\ref{figure11}(c) shows that the strengths of the magnetic field at two peaks of the blue curve are 31 G and 32 G, respectively, while the two peaks at the green curve correspond to 28~G and 26~G, respectively. This indicates that the strength of the magnetic field within the MFR decreases slightly after relaxation. In addition, the radius of the MFR  outlined by   the green curve in Figure ~\ref{figure11}(c) is slightly larger than that  outlined by the blue curve, which also means that the MFR has a lateral expansion during relaxation. Two peaks also exist on both the blue and the green curves for the electric current intensity as displayed in Figure ~\ref{figure11}(d). We notice that on each curve, the electric current intensity at two peaks are almost the same, and that the peak values of the current intensity before and after the relaxation are 3.33 $\times$ $10^{10}$ A and 2.85 $\times$ $10^{10}$ A, respectively. This indicates that the current within the MFR  weakens  after relaxation. Moreover, we also realize that the distribution of the current intensity within the MFR  tends to be more smooth and monotonous after relaxation, and the \textit{hollow core} distribution before relaxation disappears. In addition,  the feature that the overall distribution of the current  is concentrated to the inner edge of the MFR does not change.

With the global coronal magnetic configuration including an MFR constructed, we are able to locate the magnetic dip in the configuration. A magnetic dip is believed to be the place where the filament plasma is supposed to concentrate  \citep{1974A&A....31..189K}. The standard definition of a magnetic dip is the place where $B_z$ = 0 and ($\boldsymbol{B}$ $\cdot$ $\nabla$)$B_z$ \textgreater  \ 0 \citep{1993A&A...276..564T}. For a filament model that is constructed numerically, however, it is hard, if not impossible, to pin-point the location where $B_{z}$ exactly vanishes because of numerical errors  \citep{2020A&A...637A...3M}. Therefore,   we  replace this strict condition with $|B_{z}| < 0.01$. Then we look for the places where $|B_{z}| < 0.01$ in the relaxed NLFFF model and eventually, the locations of magnetic dips could be discovered at the location where  ($\boldsymbol{B}$ $\cdot$ $\nabla$)$B_z$ \textgreater  \ 0 is also satisfied.

The distribution of magnetic dips that are found this way is shown in Figure ~\ref{figure12}.  Figure ~\ref{figure12}(a) shows the top view of the dip distribution around the MFR, Figure ~\ref{figure12}(b) gives a composite of this distribution and the associated AIA~304 \AA \ image, Figures ~\ref{figure12}(c) and ~\ref{figure12}(d) show the other two side views of the same dip distribution. The AIA~304 \AA \ image in Figure ~\ref{figure12}(b) was obtained at 23:36 UT on 2015 November 14. It can be clearly seen that the distribution of magnetic  dips  basically outlines the observed filament with many important features being well duplicated although tiny deviations exist between the distribution of dips and the spine around the middle part of the filament. Important features include the overall horse-shoe-like shape of the spine, most local features of the spine, as well as distribution and extension of the left bearing barbs (see those red arrows in Figure ~\ref{figure12}(a)). The deviation of the deduced dip distribution from the observation at the north corner of the calculating domain is due to the effects  of the small-scale active region AR12453, which cannot be included in a static model. Furthermore, we notice that the dip distributes as well in the region below the MFR axis \citep{2004ApJ...612..519V}.

In addition, an interesting question is whether magnetic dips are already present to some extent in the purely potential magnetic field derived from the observed magnetogram. To answer this question, we searched them in the purely potential field. 
The result indicates that they do exist in the purely potential field as suggested by \citet{1996A&A...308..233B} although they are not as pronounced as those in the corresponding MFR configuration. This implies that the dips in the potential field are relevant to those existing in the modelled MFR configuration, which may provide a clue to the origin of the horse-shoe-like MFR and filament. But this issue is beyond the scope of the present work, and will be looked into carefully in the future.

The quasi-separatrix layer (QSL) is another crucial structure in a complex coronal magnetic configuration. It is a thin layer across which the topological  connectivity of magnetic structure changes dramatically \citep{1995JGR...10023443P, 1996JGR...101.7631D, 1996A&A...308..643D}. Such a change could be measured by the magnetic squashing degree \textit{Q} proposed by \citet{2002JGRA..107.1164T}. Three different methods to compute \textit{Q} have been suggested and compared  by \citet{2012A&A...541A..78P}. Here, we compute \textit{Q} within the 3D magnetic domain using method 3 proposed by \citet{2012A&A...541A..78P} and implemented by \citet{2015ApJ...806..171Y}. The 3D distribution of \textit{Q} is presented in Figure ~\ref{figure13}, where large \textit{Q} values (\textit{Q} $\gg$ 2) indicate the location of QSLs.   Since a 3D scenario of QSLs is usually difficult to visualize directly, we highlight the \textit{Q} values around the MFR by setting different types of colorful transparency in Figure ~\ref{figure13}. The top view of the distribution of \textit{Q} in the region nearby MFR is shown in Figure ~\ref{figure13}(a), which displays the horse-shoe-like feature clearly. In Figure ~\ref{figure13}(b), we put magnetic field lines of the MFR and the same QSL  together. Figures ~\ref{figure13}(c) and ~\ref{figure13}(d) show the other two side views of the same structures in Figure ~\ref{figure13}(b). Information revealed by comparing details of the magnetic configuration displayed in each panel in Figure  ~\ref{figure13} indicates that the filament or MFR is wrapped by a 3D QSL, but the QSL is not uniform. The component of QSL near the middle of the filament is apparently wider than those near the footpoints of the filament, which implies the existence of an area with large shearing of magnetic field around the middle of the filament. Magnetic reconnection may take place in this area, which constitutes the main cause for the relaxation \citep{2004ApJ...612..519V}.

The internal feature of QSL could be further revealed by showing the QSL distribution on the given plane. In the present work, we choose two planes of $X=-154.1^{\prime\prime}$ and $Y=-292.1^{\prime\prime}$. Figures ~\ref{figure14}(a) and  ~\ref{figure14}(b) display the global structure of MFR and the corresponding QSL slices on the planes $X=-154.1^{\prime\prime}$ and $Y=-292.1^{\prime\prime}$, respectively; and Figures ~\ref{figure14}(c) and ~\ref{figure14}(d) show QSL slices only. Figures  ~\ref{figure14}(a) and  ~\ref{figure14}(b) confirm that the MFR is indeed located inside the QSL. From
 Figures ~\ref{figure14}(c) and ~\ref{figure14}(d), we notice  that  QSL  exhibits complex internal structures, consisting of different layers  separated by different high \textit{Q} boundaries, and that the QSL extends to a relatively low altitude. In addition,  we do not find the so-called   hyperbolic flux tube (HFT), a special structure where two QSLs intersect, which  usually appears in a quadrupolar  configuration \citep{2002JGRA..107.1164T}, instead of a dipolar region.

Since the MFR is surrounded by the QSL that is the region where a significant change in the magnetic topological connectivity occurs, we take the inner edge of QSL shown in Figure ~\ref{figure14} as the boundary of MFR. This leads to the magnetic flux inside the relaxed MFR \textit{F}  = 2.26 $\times$ $10^{20}$ Mx , which is larger than the initial value of 1.82 $\times$ $10^{20}$ Mx before the relaxation, but is very close to the value of $2\times 10^{20}$~Mx given by \citet{2010ApJ...721..901S} and \citet{2012ApJ...757..168S}.  On the other hand, \citet{2008ApJ...672.1209B} apparently got a much larger value, say $1.4\times 10^{21}$~Mx. Furthermore,  the twist  and the writhe  numbers of the relaxed MFR could be evaluated by using the method of \citet{2006JPhA...39.8321B}, and we have on average  4.6 turns and 0.3 turn, respectively.  This indicates that the MFR is highly twisted. The existence of highly twisted MFR  has also been reported by  \citet{2006ApJ...645..732B}, \citet{2017NatCo...8.1330W}, \citet{2020A&A...637A...3M},  \citet{2021ApJ...917...81G} and \citet{2021ApJ...906...45D}.

The complexity of the magnetic configuration can also be described by the magnetic helicity of the system \citep{1984JFM...147..133B} and it plays an important role in the formation of the axial component of the magnetic field inside the filament, and/or in the filament channel \citep{2003SoPh..216..121M}.  So, it is worth looking into the magnetic helicity associated with the magnetic configuration including the MFR in the present work. As we showed earlier that the writhe of the MFR axis  is only 0.3,  the contribution to the MFR helicity from the writhe is negligible. In addition,  the modelled MFR is a single flux tube, in which we only consider the contribution of twist helicity. Thus the mutual helicity between the MFR and the surrounding magnetic field is also omitted. Therefore, it is appropriate that we use the twist number method \citep{2017ApJ...840...40G} to estimate the magnetic helicity.

The twist number method gives the magnetic helicity of an  MFR via estimating the twist and the axial magnetic flux. Namely, the magnetic helicity, $\mathcal{H}$ = $\mathcal{T}$\textit{F}$^{2}$,  where $\mathcal{T}$ is the average twist of the MFR and \textit{F} is the magnetic flux within the QSL that wraps the MFR. According to our results obtained earlier,  $\mathcal{T}$ = 4.6 turns  and \textit{F} = 2.26 $\times$ $10^{20}$ Mx, respectively, which gives $\mathcal{H}$ = 2.35 $\times$ $10^{41}$  Mx$^{2}$. This is consistent with  that of \citet{2018ApJ...867L...5L} but lower than that of \citet{2021ApJ...922...41T}, which found $\mathcal{H}$ = 6.8 $\times$ $10^{42}$  Mx$^{2}$. The difference may be reasonable because \citet{2018ApJ...867L...5L} and we use the same method that is different from the method used by \citet{2021ApJ...922...41T} to estimate the helicity.

\section{Discussions}
\label{sec:Discussions}

Detailed properties and fine structures of the magnetic field that supports the cool and dense solar filament/prominence in the hot and tenuous solar corona have been an open question for long time. So far, numerious efforts have been invested in this area, and many models have been constructed for the filament of medium or small scales in the region with relatively strong background magnetic field. On the other hand, for the QF in a region of weak magnetic field or large-scale IF,  very few models (e.g. \citealt{2000ApJ...543..447A}; \citealt{2004ApJ...612..519V}; \citealt{2012ApJ...757..168S}; \citealt{2020A&A...637A...3M}) have been constructed to our knowledge.

We investigated a large-scale horse-shoe-like IF that stably existed for at least 10 days. Both ground-based and space-borne telescopes have observed it. Because of the complex configuration of the filament and the weak background field in the region where the filament was located, however, the frequently used methods/approaches of extrapolation/construction of filaments do not seem to be suitable for the case we are studying here. Therefore, we adopt an improved version of the original MFR insertion method of \cite{2004ApJ...612..519V} and \cite{2007JASTP..69...24V}, namely the so-called MFR embedding method \citep{2014ApJ...790..163T}, to construct the magnetic structure of this large-scale horse-shoe-like filament. Comparing with the original MFR insertion method, the MFR embedding method possesses several advantages. First,  the magnetic field of the MFR is given in purely analytical form, which is easier to manipulate than the numerical form. Second, the axial and azimuthal flux of the MFR are estimated based on the ambient magnetic field, which seems more consistent. Third, the resulting MFR is approximately in an NLFFF  equilibrium, rather than requiring  multiple trials and iterations to reach the equilibrium as realized by the MFR insertion method.

However, only the MFR embedding method \citep{2014ApJ...790..163T} cannot fully solve our problem. There are three difficulties. First, the MFR embedding method  is not suitable for targets with arbitrary shapes. Second, it is difficult to accurately determine the path of the MFR with the method of delineating the PIL since the main structure of the filament studied here does not completely match the shape of PIL. Third,  the method of triangulation for measuring the path of filament to determine the location of the MFR in the model  (e.g. \citealt{2019ApJ...884L...1G}) does not work here as a result of the lack of observational data from multiple viewpoints.

After carefully studying and analyzing the characteristics and nature of these problems, we solved them one by one. For the first problem, we use the RBSL method developed by \citet{2018ApJ...852L..21T}, which removes the  geometrical restriction on the MFR and is suitable for MFR  of  arbitrary shape. For the second problem, we use the information provided by the AIA 304  \AA \ image to determine the path of the MFR required for the model. For the third  problem, we utilize an important technique that differs from other works. We 
started with measuring the coordinates $(X, Y)$ of the filament path according to the SDO/AIA image in 304~\AA \ that displays the filament from only one viewpoint. Then, we measure the height of the filament from the solar surface as the filament  was above the east limb of the Sun, and use this height as the $Z$ coordinate of the filament path. We note here that there is a deviation of the value of $Z$ determined this way from the true value of $Z$. But this is the best we can do for the time being. Following the steps described above, we eventually solved the three problems.

As mentioned before, constructing the MFR model with RBSL needs to deal with four free parameters, which are the filament path $C$,  the radius of the MFR cross section, $a$, the magnetic flux through the MFR cross-section, $F$, and the total current flowing through the MFR, $I$. To our knowledge, the way dealing with these free parameters in all the existing models of MFR more or less suffers from arbitrary factors. To suppress the impact of these factors, the four parameters were totally determined according to observations, which allows us to obtain a more realistic filament configuration.

With these four parameters determined, we are able to use the RBSL method to construct the MFR structure, and embed the MFR into the coronal background field extrapolated on the basis of the observed magnetogram. Then, the obtained overall magnetic field configuration (see Figure ~\ref{figure8}) is further relaxed to the best NLFFF state via a magnetofrictional iteration/relaxation. The results indicate that both the force-free and the divergence-free conditions in the final configuration are well satisfied (see Figure ~\ref{figure9}).

So far, the strength of the magnetic field in and around the solar filaments is still under debate. Table ~\ref{tab1} lists the results of previous works for the range and the average of the magnetic field related to the solar filaments. The information in  Table ~\ref{tab1} shows that the magnetic field intensity of quiescent filament/prominence is in the range of a few to tens of Gaussians, and the corresponding mean value also varies among various events. 
As for the strength of the magnetic field within and around the filament studied here, the information and the data we have collected yield a range from 7~G to 34~G, and an average of 23~G. We notice that the range of the magnetic field strength deduced here is consistent with those of \cite{1983SoPh...89....3A} and \cite{1990LNP...363...49K}, and that the average strength is close to that of \cite{2003ApJ...598L..67C}.

\begin{table*}
\begin{center}
\caption{Reported magnetic field strength of prominence/filament}

\begin{tabular}{|c|c|c|c|}

\hline 
Reference    & Magnetic Field Strength (G)   & Average  Field Strength (G)   & Type \\
\hline
\citealt{1983SoPh...83..135L}    &5$\sim$20    & 8    &Quiescent Prominence  \\

   \hline
\citealt{1983SoPh...89....3A}    &3$\sim$30    & 15   &Quiescent Prominence   \\

   \hline
\citealt{1985SoPh...96..277Q}    &6$\sim$60    & 24   &Quiescent Prominence   \\

   \hline
\citealt{1990LNP...363...49K}    &3$\sim$30    & 15   &Quiescent Prominence   \\

   \hline
\citealt{1994SoPh..154..231B}    &2$\sim$15    & 8   &Quiescent Prominence   \\

   \hline
\citealt{2003ApJ...598L..67C}    &5$\sim$70    & 20   &Quiescent Prominence   \\

   \hline
\citealt{2014AA...569A..85S}    &5$\sim$18    & 9    &Quiescent Prominence  \\
   \hline
 This work   &7$\sim$34    & 23   & Intermediate Filament   \\

\hline
\end{tabular}

\label{tab1}
\end{center}
\end{table*}

We also notice that the distribution of the magnetic field strength within the MFR is not strictly symmetric about its axis (see Figure ~\ref{figure11}(c)). Specifically, the field around its  inner edge is stronger than that around its outer edge. This feature of the magnetic field strength distribution within the MFR is consistent with the observational results of \cite{2014AA...569A..85S} and the theoretical results of \citet{2020A&A...637A...3M}. We suggest that such a feature of the magnetic field distribution inside the MFR is caused by the curvature of the MFR  \citep{1998ApJ...504.1006L}. In addition, we also find that a faster variation of the magnetic field in the transition region exists between the MFR and the background field.

Our result shows that the maximum height of the MFR  is about 78.5 Mm, and the average height is about 55.6 Mm, compared to the height of most filaments observed between 10 Mm and 100 Mm (see  \citealt{2000ApJ...543..447A} and \citealt{1990LNP...363...49K}). Comparing  with the height of ordinary QF and IF, the height of the horse-shoe-like filament studied in this work is considered low, which means that the region where the filament is located has a relatively strong background  field that imposes a strong constraint on the MFR (see \citealt{2003ApJ...588L..45R}), and helps the filament stay stable relatively long.

Analyzing the distribution of magnetic dips, especially the left-bearing barbs, of the MFR reveals the consistency with observations in H$\alpha$. This indicates that the barbs of the filament are produced by the deformation of the MFR. In the present case, we suggest that the deformation of the MFR is due to the negative fluxes located right below the barbs, and the deformation is not the kink one. Similar results have been reported by \citet{1999A&A...342..867A} and \cite{2004ApJ...612..519V}. But \citet{2017MNRAS.472.1753F} suggested that the deformation of the MFR was caused by a pair of magnetic elements of opposite polarity in the network.

At the same time, we notice that the constructed magnetic structure of the filament is a single MFR structure, only two footpoints of the MFR are anchored to the photosphere. Therefore, we conclude that the barbs produced by the distortion of the MFR are not anchored to the photosphere. The most recent observations of \citet{2020ApJ...894...64O} found that the evidence exists indeed that the barb of a filament/prominence is not anchored to the photosphere. The appearance of barbs without any parasitic polarity in the underlying magnetic distribution was also reported in the first 3D formation model of a realistic prominence by \citet{2014ApJ...792L..38X}.

This is different from the conclusion of \citet{1994ASIC..433..339M}, who suggested that the barbs extending laterally from the global structure of the filament are anchored to the photosphere, and that the fibral structure of barbs is located in the parasitic polarities below the filament. \citet{2001ApJ...560..456W} compared the 304 \AA  \ image with the magnetogram of the Michelson Doppler Imager (MDI), and found that some barbs were located in the area where the magnetic elements of opposite polarities were in mutual contact, and  that some barbs were located in the mutual contact area between the parasitic polarity and the main polarity.

However, \citet{2001ApJ...560..456W} was unable to determine the specific magnetic field polarity in the area where the barbs were located because of  the  low resolution. In this regard, we suggest that the magnetic field configuration of the filament may be a factor affecting the formation of barbs in the filament. For the filament  with an MFR configuration, the filament barbs are a natural consequence of the MFR deformation and are not anchored to the photosphere. For the filament with a sheared arcade magnetic configuration, on the other hand, the filament barbs could be anchored to the photosphere (e.g. \citealt{1994ASIC..433..339M} and \citealt{2001ApJ...560..456W}). More high-resolution observations are needed to identify and discriminate the rooting of barbs in different magnetic field configurations.

In addition, through further analyses of the distribution of  magnetic dips, we find that the magnetic dips are basically below the axis of the MFR, and are coincident well in space with the observed features of the filament in H$\alpha$. The image of the filament in 304~\AA \ indicates a relatively smooth structure with roughly uniform distribution of plasma. This suggests that the cold and dense plasma mainly concentrates in the lower area of the filament, and the hot and tenuous plasma fills the upper volume (e.g., see also \citealt{1998SoPh..183...91W}).

As we expect, the MFR structure of the horse-shoe-like filament is wrapped by a QSL with complex internal structures. We use the QSL to redefine the geometric boundary of the MFR, and calculate the axial flux in the MFR which is  2.26 $\times$ $10^{20}$ Mx, close to 2 $\times$ $10^{20}$ Mx given by \citet{2012ApJ...757..168S} for a large-scale quiet filament. Unlike the  results of \citet{2012ApJ...757..168S} that were obtained via several rounds of adjusting the initial configuration in an iterative fashion, our results were  obtained using a single relaxation step that readily relaxed to a topology close to the observations. Furthermore, we got the twist number of 4.6 turns in the  MFR on average, which is higher than that of  \citet{2008ApJ...672.1209B} and \citet{2012ApJ...757..168S} , and close to that reported by \citet{2006ApJ...645..732B} and \citet{2021ApJ...917...81G}, but lower than the value of 6  given by \citet{2021ApJ...906...45D}. All of these indicate that a highly twisted MFR may be exist in real situations.

In addition to the above progress made  on the basis of previous works, the technique for constructing the MFR  developed in this work still has room for further improvement on  three aspects. First, when determining the coordinates $(X, Y, Z)$ of the MFR path, we took the height measured when the filament appeared above the east limb of the Sun as the value of $Z$ with an implicit  assumption that  the filament height does not change during its passage of the solar disk. Second, the values of $X$ and $Y$  were  determined according to the  SDO/AIA 304 \AA  \ image of single-viewpoint only.  Third, the model developed here is a static one, which cannot deal with the configuration that is in apparent evolution. This accounts for the deviation of the deduced dip distribution from the observation at the north corner of the calculating domain where new magnetic flux was emerging.

\section{Conclusions}
\label{sec:Conclusions}

In this work, we successfully constructed the magnetic field structure for a large-scale, horse-shoe-like filament. This filament is located at the edge of a decaying active region and its background magnetic field is weak and diffuse. Hence, we combined the MFR embedding method \citep{2014ApJ...790..163T} with the RBSL method \citep{2018ApJ...852L..21T} to obtain the magnetic  structure of the  filament that is in good agreement with  observations. The main results of this work are summarized below.

(1) Three regions with magnetic structures of different topological connection exist around the filament. The kernel region nearest to the filament possesses the strongest magnetic field, the outermost region has the weakest field, and the field in the middle region has an intermediate strength.

(2) Distributions of both the magnetic field and the electric current inside the filament are asymmetric such that the magnetic field manifests large gradient in the transition region between the MFR and the background field, the magnetic field and the electric current are stronger near the inner edge than those near the outer edge of the filament.

(3) The barbs appearing in the lower part of the filament are a natural consequence of the deformation of the MFR, and are not anchored to the photosphere. The cold plasma is usually concentrated in the barb, which suggests that barbs are only seen in the wavelength of low temperature.

(4) Similar to the barbs, magnetic dips in the configuration appear also in the lower part of the filament with the distribution matching to the features observed in H$\alpha$ well. Different from barbs and dips, features observed in 304~\AA \ tend to fill the upper part of the filament. This outlines a picture of the filament with the hot plasma located higher up in the corona than the cold plasma.

\section*{Acknowledgements}

We are very grateful for the referee's valuable comments and helpful suggestions that impove the paper. This work was supported by the Strategic Priority Research Programme of the Chinese Academy of Sciences with grant
XDA17040507,  the NSFC grant 11933009,  grants associated with the Yunling Scholar Project of the Yunnan Province, the Yunnan Province Scientist Workshop of Solar Physics. The work of Y.G. was supported by NSFC grants 11773016 and 11961131002, and the MOST grant 2020YFC2201201. R.K. was supported by FWO grant G0B4521N and received funding from the European Research Council (ERC) under the European Union’s Horizon 2020 research and innovation programme (grant agreement no. 833251 PROMINENT ERC-ADG 2018). R.K. is further supported by Internal funds KU Leuven, through the project C14/19/089 TRACESpace.
The HMI and AIA data used here are avilabile on courtesy of NASA/SDO and the AIA and HMI science teams. The SECCHI data are available on courtesy of the STEREO and SECCHI teams. The ONSET data are available on courtesy of the ONSET team. We are grateful to  the  SDO/AIA, SDO/HMI, STEREO/SECCHI and ONSET consortium for the free access to the data. We thank NASA/ADS team for the convenient literature search service very much. The numerical computation in this paper was carried out on the computing facilities of the Computaional Solar Physics Laboratory of Yunnan Observatories (CosPLYO).

\section*{Data Availability}

The SDO data used in this study can be freely requested from the official website of the Joint Science Operations Center \url{http://jsoc.stanford.edu/} using the information given in the main text. The STEREO data can be freely downloaded from the website \url{https://stereo-ssc.nascom.nasa.gov/}  or via the Virtual Solar Observatory. The used ONSET data can be freely got from the ONSET team by visiting the website  \url{https://ssdc.nju.edu.cn/observationData}.



\bibliographystyle{mnras}
\bibliography{filament} 







\bsp	
\label{lastpage}
\end{document}